\documentclass[a4paper,11pt]{article}
\usepackage{pos}
\usepackage{graphicx}
\graphicspath{ {images/} }
\usepackage{wrapfig}
\usepackage{lineno}
\title{Time Calibration of the Baikal-GVD Neutrino Telescope with Atmospheric Muons}
\author[a]{V.M.~Aynutdinov}
\author[b]{V.A.~Allakhverdyan}
\author[a]{A.D.~Avrorin}
\author[a]{A.V.~Avrorin}
\author[c,d]{Z.~Barda\v{c}ov\'{a}}
\author[b]{I.A.~Belolaptikov}
\author[a]{E.A.~Bondarev}
\author[b]{I.V.~Borina}
\author[e]{N.M.~Budnev}
\author[l]{V.A.~Chadymov}
\author[f]{A.S.~Chepurnov}
\author[b,g]{V.Y.~Dik}
\author[a]{G.V.~Domogatsky}
\author[a]{A.A.~Doroshenko}
\author[c]{R.~Dvornick\'{y}}
\author[e]{A.N.~Dyachok}
\author[a]{Zh.-A.M.~Dzhilkibaev}
\author[c,d]{E.~Eckerov\'{a}}
\author[b]{T.V.~Elzhov}
\author[d]{L.~Fajt}
\author[l]{V.N. Fomin}
\author[e]{A.R.~Gafarov}
\author[a]{K.V.~Golubkov}
\author[b]{N.S.~Gorshkov}
\author[e]{T.I.~Gress}
\author[h]{K.G.~Kebkal}
\author[a]{I.V.~Kharuk}
\author[b]{E.V.~Khramov}
\author[b]{M.M.~Kolbin}
\author[i]{S.O.~Koligaev}
\author[b]{K.V.~Konischev}
\author[b]{A.V.~Korobchenko}
\author[a]{A.P.~Koshechkin}
\author[f]{V.A.~Kozhin}
\author[b]{M.V.~Kruglov}
\author[j]{V.F.~Kulepov}
\author[e]{Y.E.~Lemeshev}
\author[a,\dagger]{M.B.~Milenin}
\author[e]{R.R.~Mirgazov}
\author[b]{D.V.~Naumov}
\author[f]{A.S.~Nikolaev}
\author[a]{D.P.~Petukhov}
\author[b]{E.N.~Pliskovsky}
\author[k]{M.I.~Rozanov}
\author[e]{E.V.~Ryabov}
\author[a]{G.B.~Safronov}
\author*[b,g]{D.~Seitova}
\author[b]{B.A.~Shaybonov}
\author[a]{M.D.~Shelepov}
\author[a]{S.D.~Shilkin}
\author[f]{E.V.~Shirokov}
\author[c,d]{F.~\v{S}imkovic}
\author[b]{A.E.~Sirenko}
\author[f]{A.V.~Skurikhin}
\author[b]{A.G.~Solovjev}
\author[b]{M.N.~Sorokovikov}
\author[d]{I.~\v{S}tekl}
\author[a]{A.P.~Stromakov}
\author[a]{O.V.~Suvorova}
\author[e]{V.A.~Tabolenko}
\author[b]{B.B.~Ulzutuev}
\author[b]{Y.V.~Yablokova}
\author[a]{D.N.~Zaborov}
\author[b]{S.I.~Zavyalov}
\author[b]{D.Y.~Zvezdov}

\affiliation[a]{Institute for Nuclear Research, Russian Academy of Sciences, Moscow, 117312, Russia}
\affiliation[b]{Joint Institute for Nuclear Research, Dubna, 141980, Russia}
\affiliation[c]{Comenius University, 81499 Bratislava, Slovakia}
\affiliation[d]{Czech Technical University in Prague, Institute of Experimental and Applied Physics, 11000 Prague, Czech Republic}
\affiliation[e]{Irkutsk State University, Irkutsk, 664003, Russia}
\affiliation[f]{Skobeltsyn Institute of Nuclear Physics, Moscow State University, Moscow, 119991, Russia}
\affiliation[g]{Institute of Nuclear Physics of the Ministry of Energy of the Republic of Kazakhstan, Almaty, 050032, Kazakhstan}
\affiliation[h]{LATENA, St. Petersburg, 199106, Russia}
\affiliation[i]{INFRAD, Dubna, 141981, Russia}
\affiliation[j]{Nizhny Novgorod State Technical University, Nizhny Novgorod, 603950, Russia}
\affiliation[k]{St.~Petersburg State Marine Technical University, St.~Petersburg, 190008, Russia}
\affiliation[l]{Moscow, free researcher}

\note[\textdagger]{Deceased.}
\emailAdd{diana.seitova.18@gmail.com}

\ConferenceLogo{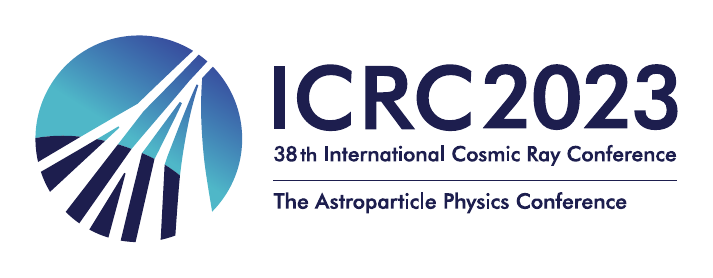}

\FullConference{
38th International Cosmic Ray Conference (ICRC2023)\\
 26 July—3 August 2023\\
  Nagoya, Japan}

\begin{document}
\maketitle
{\bf Abstract. }We present a new procedure for time calibration of the Baikal-GVD neutrino telescope. The track reconstruction quality depends on accurate measurements of arrival times of Cherenkov photons. Therefore, it is crucial to achieve a high precision in time calibration. For that purpose, in addition to other calibration methods, we employ a new procedure using atmospheric muons reconstructed in a single-cluster mode. The method is based on iterative determination of effective time offsets for each optical module. This paper focuses on the results of the iterative reconstruction procedure with time offsets from the previous iteration and the verification of the method developed. The theoretical muon calibration precision is estimated to be around $1.5-1.6~ns$.

\section{Introduction}
\begin{wrapfigure}{l}{0.6\textwidth}
\begin{center}
    \includegraphics[width=0.58\textwidth]{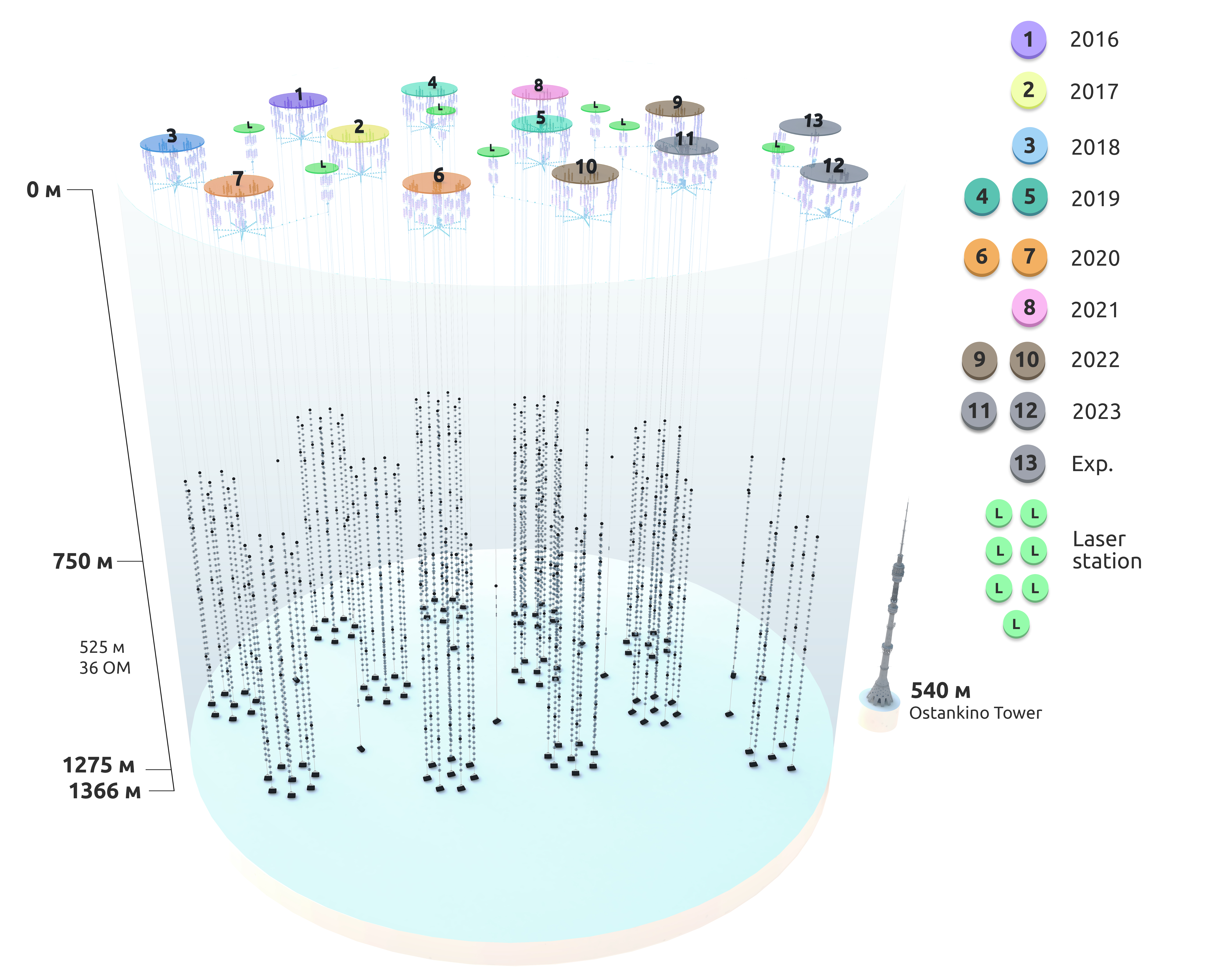}
\caption{Schematic view of Baikal-GVD neutrino telescope after 2023 winter expedition}
\label{gvd}
\end{center}
\end{wrapfigure}

Baikal-GVD is a cubic kilometre underwater neutrino telescope in Lake Baikal, currently under
construction \cite{status}. The telescope detects Cherenkov radiation from neutrino-induced secondary particles with photomultipliers (PMTs) arranged into independently operating units, clusters. At present, the telescope contains 12 clusters with 3456 optical modules (OMs) — pressure-resistant glass spheres with PMTs inside. The schematic view of the telescope is illustrated in Fig.~\ref{gvd}.

Baikal-GVD aims at detecting neutrinos of cosmic origin to study their properties, production mechanisms and sources. The search for astrophysical neutrino sources is performed via reconstructing the direction and energy of the particles detected. For the point-source search, it is crucial to get a good directional resolution of the neutrino so as to achieve a good performance in the reconstruction procedure. In order to achieve the required performance of the telescope, accurate timing measurements
(order of $1~ns$) of photons detected with PMTs are of great importance.

Atmospheric muon bundles reconstructed in a single-cluster mode are used for this purpose. The method
described here is based on the ideas previously implemented in the NT-200, AMANDA \cite{amanda} and ANTARES  \cite{antares} neutrino telescopes. 

In order to do the calibration with muons, muon tracks should be reconstructed with preliminary time calibration.  For time calibration, Baikal-GVD uses artificial light sources, such as LEDs \cite{calib} and lasers, and the achieved precision is $3-5~ns$. In this work, the LED calibration was used as a preliminary. With the new methodology described in this paper, we aim to obtain a better calibration with a $1-3~ns$ precision.

\section{Time calibration procedure}
The iterative method for obtaining time corrections for every optical module using atmospheric muons has been developed. It consists of the following steps:
\begin{itemize}
\item the subset of hits which best fits the single muon model is selected with the efficient ScanFit hit finding algorithm \cite{hitsel};
\item the hit in the optical module under study ({\it probe} hit) is removed from the subset, and the remaining {\it reco} hits are used for reconstruction of the muon track with the standard reconstruction algorithm \cite{reco}. The exclusion of {\it probe} hits guarantees that resulting time residual distributions are not biased since the muon track has not been fitted to minimize them;
\item given the muon passing near the optical module and emitting a Cherenkov photon which strikes this OM, the residual time, $t_{res}$, for the {\it probe} hit is defined to be the difference between the measured real time of the photon's arrival at the OM, $t_{meas}$, and the expected theoretical arrival time from the analysis of the single muon track model, $t_{theor}$:
$$t_{res} = t_{meas}-t_{theor}$$
and the time distribution histogram is filled with obtained $t_{res}$ values;
\item the median value is obtained from the time distribution histogram and used in the calculation of time correction for the next iteration.
\end{itemize}
%\begin{wrapfigure}{r}{0.6\textwidth}
\begin{figure}
\begin{center}
\includegraphics[width=0.65\textwidth]{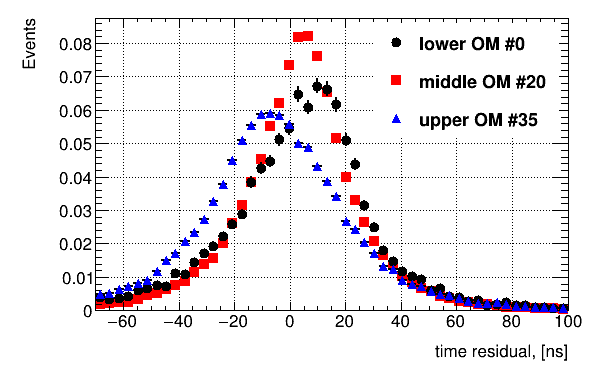}
\caption{Simulated time residual distributions for 3 Baikal-GVD optical modules: the lower one, the middle one and the upper one }
\label{tres}
\end{center}
\end{figure}
%\end{wrapfigure}

This procedure is performed individually for all optical modules. In order to minimize the uncertainties of the method, the resulting time correction, $t_{corr}$, for a given optical module is defined with respect to median values of time residual distributions obtained from Monte Carlo simulations where no miscalibrations are present ($t^{med}_{res, MC-T0}$): 
\begin{equation} \label{eqn}
    t_{corr} = t^{med}_{res, exp} - t^{med}_{res, MC-T0}.
\end{equation}
Time corrections for every optical module are determined in this way, and the above described procedure is repeated iteratively until the corrections become small and converge to some limit.  

\begin{figure}[h]
\begin{minipage}{0.49\textwidth}
  \includegraphics[width=0.95\textwidth]{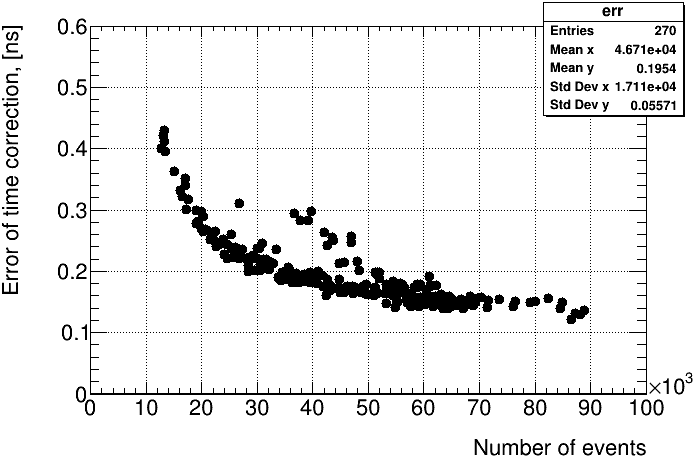}  
\end{minipage}
\begin{minipage}{0.49\textwidth}
  \includegraphics[width=0.95\textwidth]{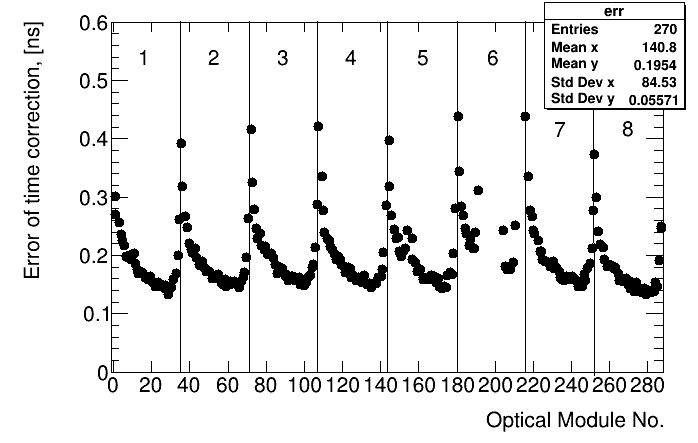} 
\end{minipage}
\caption{{\bf Left:} Statistical error of time correction defined at last iteration vs number of events. { \bf Right:} Error vs optical module No }
\label{stat} 
\end{figure}

The time residual distributions for simulated data after the first iteration are shown in Fig.~\ref{tres}. The determination of the median of time residual distributions depends on statistics, therefore, statistical errors were calculated in the same way as the error of the mean. The resulting time correction was defined according to Eq.~\ref{eqn}, therefore, the resulting time correction error was calculated as a squared sum of two errors. The dependence of the time correction statistical error on the number of events for the simulated data at the last iteration is illustrated in Fig.~\ref{stat} (left). In this plot, we see that for some optical modules the errors deviate from the common behaviour. The nature of these deviations is not clear at the moment and requires a further study. 
%with uniformly shifted hit times
 Figure~\ref{stat} (right) shows the time correction error for every optical module in one cluster.  The cluster contains 8~strings with 36~optical modules on each (in the cluster considered, 18 OMs did not operate in 2019), the numeration of the modules starts from the bottom. The figure is divided into 8~sections for a clear illustration where every section represents the corresponding string with its optical modules. The calculated error of time correction for the optical modules located closer to the lake bottom is larger since there are less atmospheric muons reaching the bottom and, consequently, less hits at lower optical modules. The errors slightly increase for upper-lying optical modules since the optical modules located at the string edge form less hit pairs compared to the middle ones during the hit selection stage. However, due to the larger atmospheric muon flux that reaches the upper modules, the errors are not as big as for lower-lying modules.  

\section{Verification of the method}
To evaluate whether the method described can reduce miscalibrations in the experiment, the algorithm was verified on the Monte Carlo simulated sample where hit detection times in every optical module were shifted/miscalibrated according to the Gaussian distribution with $\sigma = 3~ns$ and, alternatively, uniform distribution between $-15~ns$ and $+15ns$. The extent of miscalibration in the experiment is unknown, therefore, the extreme case with uniformly distributed $\pm 15~ns$ shifts is considered. 

The method described was tested using the simulated data sample of the Season 2019 from Cluster 1, which is equivalent to 30~days of the telescope exposition time. Two cases were analyzed:
\begin{itemize}
\item all reconstructed tracks with hits at least on $2$~strings and $7$~optical modules;
\item tracks with the following selection criteria to minimize the number of misreconstructed events — tracks with the visible length in the detector more than $200~m$ and with a hit in one of the adjacent optical modules to the probe hit OM.
\end{itemize}

\begin{figure}[h]
\begin{minipage}{0.49\textwidth}
  \includegraphics[width=0.95\textwidth]{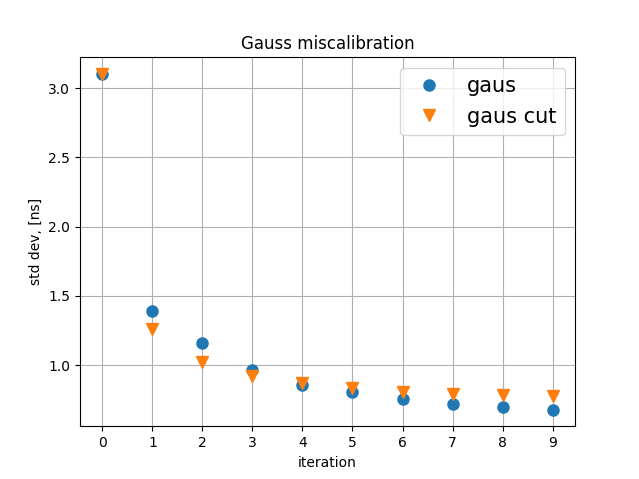}  
\end{minipage}
\begin{minipage}{0.49\textwidth}
  \includegraphics[width=0.95\textwidth]{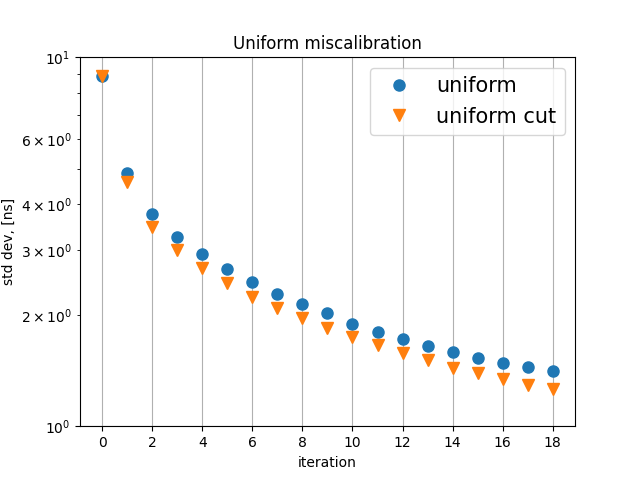} 
\end{minipage}
\caption{Width of remaining miscalibration after every iteration. Iteration 0 corresponds to initially inserted miscalibration according to Gaussian ({\bf left}) and uniform ({\bf right}) distributions. Others demonstrate how miscalibration diminishes with iteration count. }
\label{miscalib} 
\end{figure}

For two cases mentioned above, time corrections were determined for every optical module. To see whether this algorithm reduces artificially implemented miscalibrations, the change of miscalibrations after applying the time corrections derived after every iteration was analyzed. The width of the remaining miscalibrations is obtained and illustrated for Gaussian shifts in Fig. \ref{miscalib} (left) and for uniformly distributed shifts Fig.~\ref{miscalib} (right). The algorithm diminishes Gaussian shifts with $\sigma = 3~ns$ to $0.68~ns$ and $0.78~ns$ for all reconstructed muon and selected muon tracks, respectively. There were 9 iterations, and in both cases widths converge to these values. With uniformly distributed shifts, 18 iterations are not enough for convergence, and at the last iteration, the widths acquired are $1.41~ns$ and $1.26~ns$ for these two cases. 

\section{Results and discussion}
\begin{figure}[h]
\begin{minipage}{0.49\textwidth}
\begin{center}
\includegraphics[width=0.98\textwidth]{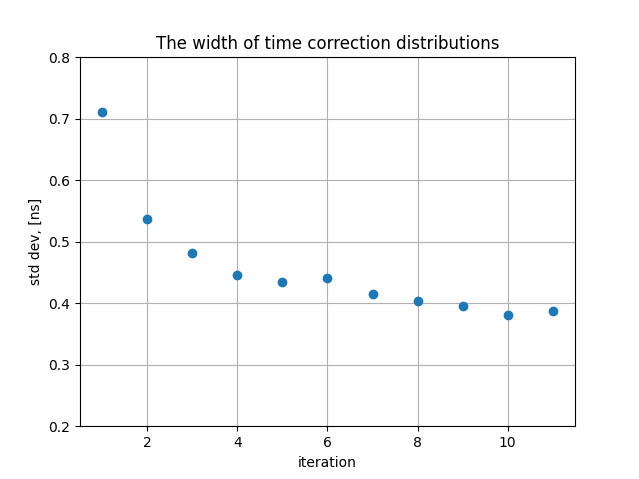}
\caption{Width of experimental time corrections obtained for Season 2019 Cluster 1  }
\label{w-exp}
\end{center}
\end{minipage}
\begin{minipage}{0.49\textwidth}
\begin{center}
\includegraphics[width=0.98\textwidth]{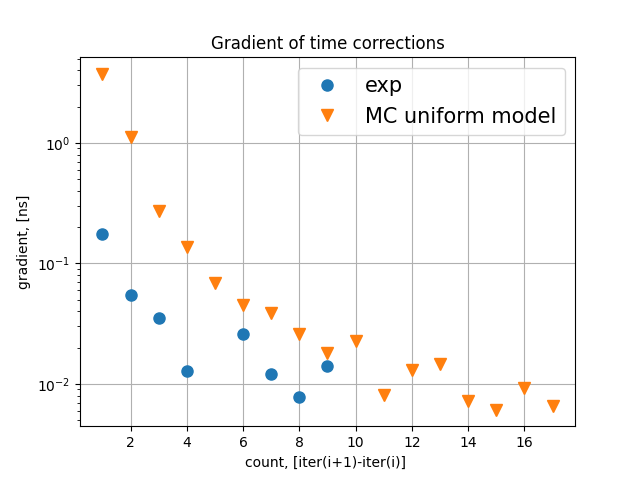}
\caption{Gradient of time corrections for experimental data and MC model, Season 2019 Cluster 1  }
\label{grad}
\end{center}
\end{minipage}
\end{figure}

The width of the distribution of the experimental time corrections measured for the Season 2019 and Cluster 1 is demonstrated in Fig.~\ref{w-exp}. The time correction values were obtained with respect to $t^{med}_{res, MC-T0}$. The width diminishes with iteration count and nearly converges to $0.4$~ns in 11~iterations. The total time correction for the data considered is defined as the sum of corrections obtained in 11~iterations.

The precision was estimated theoretically by comparing the MC model with uniformly shifted miscalibrations to the experimentally obtained results. For that purpose, we analyzed the form of the distribution of the widths of time corrections (plots similar to Fig.~\ref{w-exp}) both for the experiment and theoretical model. By comparing how the difference of widths changes from iteration to iteration (gradient of widths), we can see that the experimental gradient assimilates the theoretical model starting between the 3rd-4th iterations as illustrated in Fig.~\ref{grad}. From this analysis, we concluded that the precision of the preliminary calibration was around $2.7-3~ns$ (Fig.~\ref{miscalib} (right)). By extrapolating 11 iterations along the miscalibration plot (Fig. ~\ref{miscalib} (right)), we estimated the acquired precision to be around $1.5-1.6~ns$. The reliability of the precision estimation is still under discussion.

%\begin{wrapfigure}{r}{0.6\textwidth}
%\end{wrapfigure}

\section{Conclusions}
In this paper, we have presented the method for time calibration with reconstructed atmospheric muons. The method allows determining effective time offsets for every optical module in addition to the calibration values obtained by the earlier implemented techniques. Using muons, to achieve a $1.5-1.6~ns$ precision is possible. The method is still under development, and our further interest is in studying how the uncertainties of measurements of the optical module position (geometry) affect the acquired precision. 

The work is supported by the JINR grant for young scientists and specialists No 23-202-08. 

We would like to thank the team of the JINR cloud infrastructure for providing computing resources.

%% Full authors list (ONLY FOR COLLABORATIONS)
%\clearpage
%\section*{Full Authors List: \Coll\ Collaboration}
%
%\noindent \textbf{Note comment afterwards:} Collaborations have the possibility to provide an authors list in xml format which will be used while generating the DOI entries making the full authors list searchable in databases like Inspire HEP. \\
%
%\scriptsize
%\noindent
%first.author$^1$, 
%second.author$^2$, 
%third.author$^3$ % .... more names
%and 
%last.author$^{n}$ \\
%
%\noindent
%$^1$first.affiliation.
%$^2$second.affiliation. % .... more affiliation
%$^{m}$last.affiliation.


\begin{thebibliography}{99}
\bibitem{status} The Status of Baikal-GVD, these proceedings.
\bibitem{amanda} AMANDA Collaboration, Time Calibration of the AMANDA neutrino telescope with cosmic ray muons, Proceedings of ICRC 2001.
\bibitem{antares} ANTARES Collaboration, Time Calibration with atmospheric muon tracks in the ANTARES neutrino telescope, arXiv:1507.04182v1
\bibitem{calib} Baikal-GVD Collaboration, Time calibration of the neutrino telescope Baikal-GVD, EPJ Web of Conferences 207, 07003 (2019).
\bibitem{hitsel} Baikal-GVD Collaboration, An efficient hit finding algorithm for Baikal-GVD muon reconstruction, arXiv:2108.00208 [astro-ph.IM]
\bibitem{reco} Baikal-GVD Collaboration, Performance of the muon track reconstruction with the
Baikal-GVD neutrino telescope, arXiv:2108.11217 [astro-ph.HE]


\end{thebibliography}
\end{document}